A Prospective ISRO-CfA Himalayan Sub-Millimeter Observatory Initiative

T. K. Sridharan
Center for Astrophysics | Harvard and Smithsonian

Contributors:

Shmuel Bialy
Ray Blundell
Andrew Burkhardt
Tom Dame
Sheperd Doeleman
Douglas Finkbeiner
Alyssa Goodman
Paul Grimes
Nia Imara
Michael Johnson
Garrett Keating
Charles Lada
Romane Le Gal
Philip Myers
Ramesh Narayan
Scott Paine
Nimesh Patel
Alexander Raymond
Edward Tong
David Wilner
Qizhou Zhang
Catherine Zucker

# A Prospective ISRO-CfA Himalayan Sub-Millimeter Observatory Initiative

## Abstract

The Smithsonian Astrophysical Observatory (SAO), a member of the Center for Astrophysics | Harvard and Smithsonian, is engaged in discussions with the Space Applications Centre (SAC) of the Indian Space Research Organization (ISRO) and its partners in the Indian Submillimetre-wave Astronomy Alliance (ISAA), to collaborate in the construction of a submillimeter-wave astronomy observatory in the high altitude deserts of the Himalayas, initially at the 4500 m (14700 ft) Hanle site of the Indian Astronomical Observatory. The Himalayan sites are among a small number of such locations globally, with sufficient atmospheric transmission to allow observations in the sub-millimeter wavelength bands, at frequencies of about 200-500 GHz and higher. ISAA is a partnership of scientists from leading Indian astronomy, physics and space research institutions. The initiative targets two primary science goals. One is the mapping of the distribution of neutral atomic carbon, and the carbon monoxide (CO) molecule in higher energy states, in large parts of our home Galaxy, the Milky Way, and selected external galaxies. Such studies would advance our understanding of molecular hydrogen present in the interstellar medium, but partly missed by existing observations of the commonly used tracer molecule CO; and characterize Galaxy-wide molecular cloud excitation conditions, through multi-level CO observations. Stars form in interstellar clouds of molecular gas and dust, and these observations would allow research into the formation and destruction processes of such molecular clouds and the life cycle of galaxies. As the second goal, the observatory would add a new location to the global network of telescopes which operate in synchronism to synthesize the earth sized telescope called the Event Horizon Telescope (EHT), through the Very Long Baseline Interferometry (VLBI) technique. The EHT recently imaged for the first time, the signatures of a supermassive black hole at the center of the M87 galaxy. The EHT network lacks a station in the Himalayan longitudes. This addition would enhance the quality of the images synthesized by the EHT, support observations in higher sub-millimeter wave bands, sharpening its resolving ability, improve its dynamic imaging capability and add weather resilience to observing campaigns. In the broader context, this collaboration can be a starting point for a wider, mutually and scientifically beneficial engagement between the Indian and US astronomy communities, including a potential future EHT space component.

## Extended Synopsis

The Space Applications Centre (SAC) of the Indian Space Research Organization (ISRO) is developing a sub-mm wavelength astronomy observatory to be constructed and operated at the 4500 m Hanle high altitude desert site of the Indian Astronomical Observatory in the Himalayas. SAC is a founding member of the Indian Submillimetre-wave Astronomy Association (ISAA), an informal partnership of leading Indian physics, astronomy and space research institutions, newly formed to provide science support to this initiative on a national basis.

There is significant interest at the Center for Astrophysics | Harvard and Smithsonian in the science opportunities presented by this development. Given the expertise in high sensitivity sub-mm astronomy instrumentation developed at the CfA, and its pioneering applications, a mutually beneficial opportunity to participate in the Indian observatory development exists.

Accordingly, the submillimeter receiver laboratory proposes to assist in the design, development and construction of the high sensitivity superconducting receivers and antenna for the Indian observatory. In return, the opportunity to shape and pursue a science program of current and future interest to the CfA, enabled by the high altitude Himalayan sites, is presented. This also opens a path to potential future longer term developments in space.

Two broad areas of science have been identified viz., – studies of the neutral interstellar medium (ISM), atomic and molecular, and the transition between them, and the enhancement of the capabilities of the Event Horizon Telescope (EHT).

Under the ISM studies, the Indian observatory would be used to jointly map large parts of the Milky Way and selected external galaxies in the emission lines of neutral atomic carbon at 492 GHz and the higher energy rotational transition (4-3) of the carbon monoxide (CO) molecule. High, dry sites, such as Hanle and others in the Ladakh region, among a few globally, with low atmospheric absorption, are necessary for these observations. A new receiver would be developed for this purpose. The atomic carbon line emission is expected to address multiple science avenues related to the formation of molecular clouds and a poorly characterized phase of the molecular ISM missed by the currently common observational probe, CO.

Under the EHT science, the Himalayan antenna would add a station in a longitude range currently lacking coverage and support EHT operation in higher frequency bands. These developments would improve the quality and resolution of the images of super massive black holes synthesized by the EHT, enhance weather resilience of EHT observing campaigns, and add to its dynamic imaging capabilities.

The lower frequency receivers in the 230 and 345 GHz EHT bands would also carry out Galaxy-wide large scale mapping of the CO(2-1) and (3-2) emission lines. Along with the CO(4-3) mapping, these observations would allow Galaxy-wide characterization of the excitation conditions in molecular clouds for the first time, complementing and adding value to the CfA CO survey.

The details of the proposed observatory, science and instrumentation would be defined jointly by SAC, ISAA and CfA, to address the above science goals. A 3 m antenna is under development at the SAC and a 6 m antenna is under consideration, with the Smithsonian Submillimeter Array (SMA) antenna design being a candidate.

In the longer term opportunities to expand the collaboration to develop a space VLBI component to the EHT and into related areas exist and can be enabled by the proposed initial efforts.

## Executive Summary

The Indian Space Research Organization, ISRO, is the primary space agency of the Government of India, possessing capabilities spanning the full gamut of space activities, with growing emphasis and successes in space science. The Space Applications Centre (SAC) is a research center of ISRO focusing on the design and development of communication, navigation, remote sensing and science payloads. SAC's capabilities cover optical, infrared and radio wavelengths. Science instruments delivered include payloads for the Mars Orbiter Mission, Chandrayaan-1 & -2 lunar missions, synthetic aperture radars and unfurlable antennas.

SAC is currently developing sub-mm/THz engineering capacity, having built a mm-wavelength (183 GHz) water vapour sounder. This sub-mm/THz initiative is in a fully funded ground based demonstration phase, to build and operate a sub-mm wavelength astronomical observatory in the Himalayas. The Indian Sub-mm-wave Astronomy Alliance (ISAA), formed to support this initiative, brings together scientists from SAC, ISRO, Raman Research Institute (RRI), Tata Institute of Fundamental Research (TIFR), Indian Institute of Space Science and Technology-ISRO and Indian Institute of Astrophysics (IIA). An observatory with a 3 m antenna and moderate sensitivity 230/345 GHz instrumentation is under development. A larger 6 m antenna with state-of-the-art superconducting receivers for mm/sub-mm bands is under consideration.

The Himalayan high altitude desert region of the Ladakh division has numerous potential high altitude sites (5000-6000 m) accessible by partly paved roads. Such sites could provide access to sub-mm and THz wavelength bands for astronomical observations. Infrastructure exists at the 4500 m Hanle site of the Indian Astronomical Observatory of the IIA, where a 2 m optical/NIR telescope (since 2003), a TeV gamma-ray array (HAGAR, TIFR-IIA, since 2008) and the GROWTH-India 0.7 m robotic telescope (US-India:NSF/Caltech-DST/IIA,IITB; since 2018) are operational and a 21 m Cherenkov telescope (MACE; Dept. of Atomic Energy-TIFR,IIA) is expecting first light. A large national solar telescope is under design while a 10 m class optical/IR telescope is being planned. Site testing at 220 GHz shows atmospheric transmission similar to Mauna Kea with better winter months.

The Smithsonian Astrophysical Observatory (SAO), a member of the Center for Astrophysics | Harvard and Smithsonian (CfA) is a pioneer in mm/sub-mm/THz astronomy and technology. The staff of the SAO and the CfA (1) developed, deployed and currently operate the first sub-mm wave interferometric aperture synthesis telescope (the Submillimeter Array, SMA), on Mauna Kea, Hawaii (2) made the first ground-based THz astronomical observations using a telescope developed and deployed by SAO's Submillimeter Receiver Laboratory at 5,525 m altitude in northern Chile (3) carried out the largest surveys of molecular clouds, mapping the entire Milky Way Galaxy (the CfA CO survey) (4) led the formation of the Event Horizon Telescope (the EHT) which imaged the signatures of the Super Massive Black Hole in M87, winning a 2020 Breakthrough prize. These efforts were enabled by broad expertise in mm/sub-mm wavelength instrumentation and astronomy in general and in the area of high sensitivity superconducting (SIS) front end receiver technology, a field in which SAO's Submillimeter Receiver Laboratory is a leader.

As part of its sub-mm/THz initiative, ISRO is exploring high sensitivity SIS receiver technology development and has sought participation from the SAO Receiver Laboratory through collaboration. In return, opportunities to jointly shape and pursue new science from the high Himalayan sites are offered at little cost. As the overarching goal of the CfA is science, we evaluated the science prospects of the opportunity. We identified two broad areas with significant current and future possibilities and interest within the CfA community, briefly outlined below. The science opportunities in areas of interest to the CfA created by the planned Indian sub-mm observatory, and the need for sub-mm astronomy experience and SIS receiver technology expertise to enable highest sensitivity observations, provide natural avenues for collaborative work between the SAC-ISRO and the CfA.

*1. CO-dark-$H_2$ phase of the ISM*: The central process in galaxy evolution is the life cycle of the ISM - the diffuse atomic gas condenses into dense molecular ISM, followed by clump, core and star formation, leading to the destruction and dispersal of the dense phase back into the diffuse medium, through radiation, stellar winds and supernovae. The dominant component of the cold dense medium, $H_2$, is not directly observable and is usually traced by its easily observed proxy, CO. However, a component of the dense molecular ISM lacks sufficient CO – the *CO-dark-$H_2$*, due to differences in formation, shielding and dissociation by cosmic rays (CR) and far ultraviolet (FUV) radiation, compared to $H_2$. The surface regions of molecular clouds and the clumps within, and their interfaces with young stars, are dominated by *CO-dark-$H_2$* and are missed by CO observations. Therefore, the cloud and clump formation and destruction processes which operate in these regions, are not fully accessible to current observations. Theory, simulations and limited observations identify atomic carbon C as a tracer of this hidden phase and indicate differences with CO maps. As the C emission depends on metallicity and CR and FUV environments, it is necessary to make large scale maps of the Galaxy and of individual clouds, covering a range of conditions and Galactocentric radii. Observations of our nearest spiral galaxies M31, M33 which harbors a large metallicity gradient, and M51, are also possible, and important as bench marks for increasingly prevalent high-z, neutral atomic carbon (C[I]) observations. Further, C[I] detection surveys of more distant galaxies are feasible. Additionally, these observations can also inform scaling relations in molecular clouds by tracing new material; and provide an additional new template in decomposing the Galactic diffuse gamma ray emission into its components, whose residuals are used to indicate new processes e.g., dark matter annihilation/decay.  Thus, it is important to study the *CO-dark-$H_2$* component from multiple science perspectives. Dedicated large scale surveys in the C[I] fine structure line at 492 GHz are well suited to the proposed small telescope, and will expand and add value to the CfA CO Survey. Observations of this line are very limited due to the need for high, dry sites, such as in the Ladakh region, among a few globally. The development of a 461/492 GHz receiver targeting the C[I] & CO(4-3) lines is proposed, which will also provide an option for a future additional SMA band, enabling high resolution studies of immediate vicinities of young massive stars and of bright individual molecular clouds in nearby galaxies.

*2. Event Horizon Telescope, mm & sub-mm VLBI*: With the CfA-led first imaging of the signature of the supermassive blackhole in M87 achieved, the EHT now targets improvements to imaging quality and resolution, and dynamic imaging as its next steps. The path involves expanding the EHT into its next generation, with additional stations and higher frequency operation. A VLBI station in the Himalayas, a longitude range lacking current EHT coverage, enables advances on both fronts. Such a station would have mutual visibility with a subset of existing stations for some of the key targets - e.g.: M87, OJ287 and Sgr A*. Even without adding unique spatial frequency coverage, additional stations increase the likelihood of successful observations, which must be planned well in advance with limited predictability of joint weather conditions across the diverse EHT stations, and additionally provide calibration redundancies. In the context of other additional stations being pursued, a Himalayan station enhances the EHT with additional baselines and closure triangles. With the highest resolution at 230 GHz reached for earth based VLBI, higher frequency (345 GHz) operation is a critical next step, supported by the Himalayan sites.  The EHT-VLBI work would involve equipping the observatory with 230 and 345 GHz receivers and even higher frequencies in future. These receivers can also be used for large scale surveys in CO (2-1) and (3-2) and other species when the weather precludes 461/492 GHz *dark-$H_2$* observations. In the longer term, considering ISRO's space capabilities, interest, and increasing emphasis on space sciences, a future space VLBI experiment is a possibility. Initial Himalayan pursuits are well worth the effort and resonate with nascent CfA initiatives.

*3. Additional Science:* Other areas of scientific interest are (1) monitoring variability of strong spectral line targets – the emerging field of episodic accretion, flaring events marking onset of maser emission from massive young stellar objects (2) spectral line surveys towards selected targets and  (3) joint work in modeling data from SAC's atmospheric water vapour sounder to be flown by ISRO in the near future.

*Collaboration Model and Plans:* To pursue these possibilities, a Memorandum of Understanding between the SAO and the SAC is under development. Under this MoU, a phased approach to fully develop the initiative is envisaged. In the first phase, scientists and engineers from the CfA and the SAC, and also from the ISAA institutions, will work jointly to define the science goals and instrument requirements in more detail. In parallel, engineering staff from SAC will spend time at the SAO Receiver Laboratory, participating in the on-going wide-band SMA (wSMA) receiver upgrade work through a staff exchange program to be set up. This allows hands-on experience in constructing, testing and operating sub-mm receivers, while contributing technical effort to wSMA build-out. Towards the end of this period, one set of 230 & 345 GHz band receivers will be built by the SAC engineers for installation on the Himalayan antenna. Design and development work towards new 461/492 GHz receivers will be jointly pursued during the same time. A 6 m antenna meets the science requirements, and options for its design and construction, will be explored. Building a 6 m antenna of the SMA design by Indian vendors is under consideration. SAC is also working on a 3 m antenna to be developed on a shorter time scale. The second phase targets commissioning and initial science operations, envisaged at the 4500 m Hanle site. While 461/492 GHz operation can be tested and carried out at Hanle, and participation in EHT expansion can be accomplished, the full CI mapping science program will greatly benefit from a move to a higher site, in a later third phase.

In recent years, India has embarked on a number of "mega science" programs, participating in large international projects like LIGO-India, TMT, MSE, SKA, CERN, ITER, FAIR and NASA-ISRO SAR (NISAR). Locally developed examples are the Indian Neutrino Observatory, MACE (Cherenkov), GMRT, Astrosat, Chandrayaan-1 & -2 and the Mars Orbiter Mission projects. The NASA-ISRO collaboration on Chandrayaan-1 was particularly fruitful, delivering a high impact result – discovery of lunar water. Leveraging the ISRO-funded sub-mm/THz seed effort, we propose to seek additional funding and support under the framework of the US-India Joint Civil Space Working Group, the Indo-US Science and Technology Forum (IUSSTF) and the NSF-DST PIRE program. Successful precedents under Indo-US government programs are the NASA-ISRO NISAR, Chandrayaan-1 and the Growth-India (NSF/Caltech-DST/IIA, IITB) robotic telescope in the Himalayas. With this context and history, collaboration with Indian astronomy institutions and ISRO holds great potential and promises mutual benefits in a broad range of areas, with the present sub-mm initiative being the proverbial tip of the iceberg.

**Table of Contents**

**1. The Indian Space Program**

The Indian Space Research Organization, ISRO, is the primary space agency of the Government of India. Starting with a charge to deliver societal benefits from space efforts, ISRO has grown to encompass the full gamut of space activities - launch vehicles, diverse payloads, ground stations, command, control and tracking, and exploitation. With capability to deliver 4 tonnes to GTO and 8 tonnes to LEO and recent successes - e-g.: *Chandrayaan-1*, which discovered lunar water; the *Mars Orbiter* mission with successful orbit insertion at first attempt; a multiwavelength imaging X-ray/UV astronomy observatory, *AstroSat*, with X-ray polarization capability - ISRO's profile is changing with growing emphasis on space sciences. In September 2019, the *Chadrayaan-2* mission placed a suite of instruments in lunar orbit, including a dual frequency (L and S band) full polarimetric synthetic aperture radar (SAR) to probe subsurface ice, and attempted to deploy a lander and a rover to explore the lunar south polar regions for the first time. The solar probe *Aditya-L1* and the NASA-ISRO SAR (*NISAR*) target 2020 and 2021 launches. An Epoch Of Reionization (EoR) payload is under study and a probe to Venus and a second mission to Mars are in planning phases. ISRO is in the process of drawing up long range vision plans to shape its future growth, including astronomy and planetary missions. All aspects of the science programs are managed by ISRO's Space Science Program Office (SSPO) at the ISRO HQ, Bangalore.

The Space Applications Centre (SAC) is a research center of ISRO focusing on the design and development of communication, navigation, remote sensing and science payloads. SAC's capabilities cover optical, infrared and radio wavelengths. Instruments delivered include payloads for the *Mars Orbiter* mission, *Chandrayaan-1 & -2* lunar missions, synthetic aperture radars (SAR) for *Chandrayaan-2* and *NISAR* (earth science) missions and unfurlable antennas. The EoR payload being studied is a joint Raman Research Institute (RRI) – SAC project.

In a new initiative, the SAC is currently developing sub-millimeter/THz wavelength capabilities, building on prior work at mm wavelengths in developing a 183 GHz (~ 1.5mm) radiometer for an atmospheric water vapor sounder payload. This ambient temperature Schottky mixer radiometer was developed by a dedicated group – the Terahertz and Sub-millimetre Development Division (TSDD), established to pursue sub-mm/THz capabilities. The sub-mm/THz initiative is currently in a fully funded demonstration phase with a goal to design, construct and operate a sub-mm wavelength astronomical observatory in the Himalayas. A 3 m diameter meter antenna is under development along with cooled sub-harmonically pumped Schottky receivers for the 230 and 345 GHz bands. An end-to-end demonstration of this system at the 4500 m Hanle site in the Himalayas (section 2) is targeted for the summer of 2021. A larger 6 m sized antenna and higher sensitivity state-of-the-art instrumentation for sub-mm wavelength observations are under consideration.

ISRO has a history of successfully collaborating with other national academic institutions, e.g., on the *Astrosat* mission - the X-ray payloads were developed by the Tata Institute of Fundamental Research (TIFR), the Raman Institute and ISRO, the UV payloads were developed by the Indian Institute of Astrophysics (IIA) and the science user interface is managed by the Inter-University Centre for Astronomy and Astrophysics (IUCAA). The *Aditya-L1* mission to be flown in 2020 and the EoR payload under study are joint ISRO-IIA-IUCAA and ISRO-RRI projects, respectively. ISRO also collaborates with international institutions, recent examples being the two *Chandrayaan-1* instruments that found lunar water – the NASA-JPL/Brown University Moon

Minerology Mapper (M3) and the NASA Mini-SAR, the NASA lunar reflector array on the Chandrayyan-2 lander, and the aforementioned large *NISAR* earth science mission to deploy a dual band radar system with a 12 m antenna.

## 2. Astronomy Infrastructure in the Indian Himalayas

The Himalayan high altitude desert region of the Ladakh division has numerous potential high-altitude sites, accessible by partly paved roads, which could provide access to sub-mm and THz wavelength bands for astronomical observations. Indeed, infrastructure exists at the 4,500-m Hanle site of the Indian Astronomical Observatory (IAO; Prabhu, 2014) where a 2 m optical/NIR telescope (IIA, since 2003), a 0.7 m US-India robotic telescope (since 2018; Caltech-IIA, Indian Institute of Technology, Bombay: NSF/Partnerships for International Research and Education, PIRE-Department of Science and Technology, India, DST/Science and Engineering Research Board, SERB) and a TeV gamma-ray array (HAGAR, TIFR-IIA, since 2008) are operational and a large 21 m Cherenkov telescope (MACE; Dept. of Atomic Energy-TIFR-IIA) is under commissioning, expecting first light this winter. The *SARAS-2* EoR experiment (RRI) operated in this region (Singh et al. 2018) and newer versions are under consideration. A National Large Solar Telescope (NLST) is under design for construction in the Ladakh region.  A 10 m class National Large Optical/IR Telescope  (NLOT) is being planned, also to be sited in the Ladakh region. The NLOT group of IIA is collaborating with the Maunakea Spectroscopic Explorer (MSE) group to jointly develop the design of the two telescopes which are very similar (IIA 2017/18 Annual Report; India/IIA is an MSE partner).

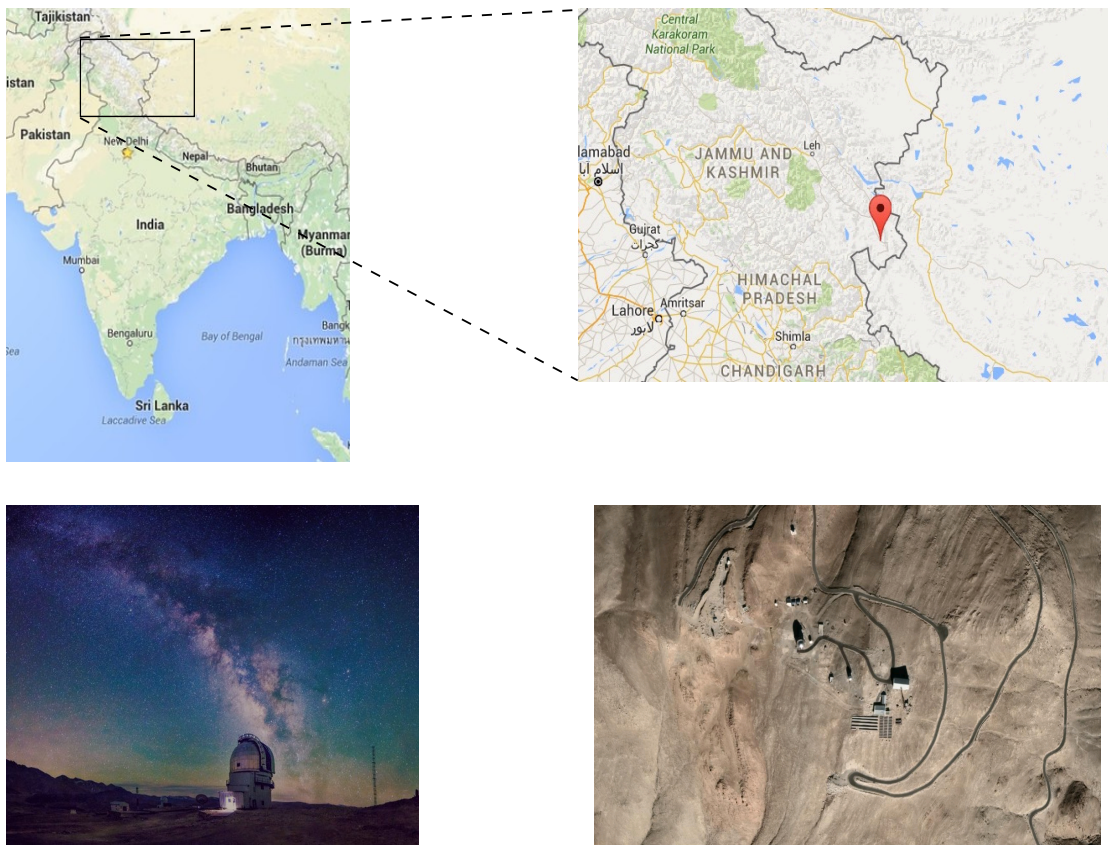

Figure 1. (a) The location of the Ladakh division of the Indian Himalayas. (b) The Hanle site of the Indian Astronomical Observatory at 75° 57' 51" E 32° 46' 46" N. (c) A satellite image of the 4500 m summit area. (d) The 0.7 m US-India (NSF/PIRE-DST/IIA, IITB) GROWTH-India robotic telescope for observing transients. (image credits: a,b,c: google; d: Navneeth Unnikrishnan)

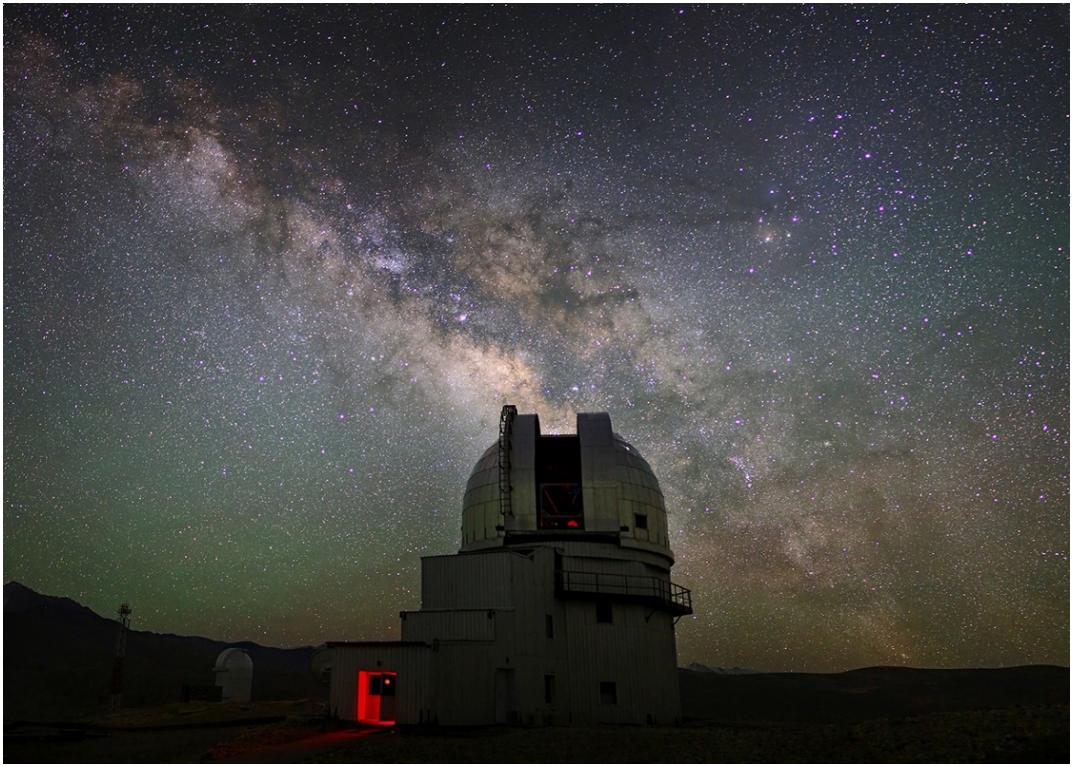

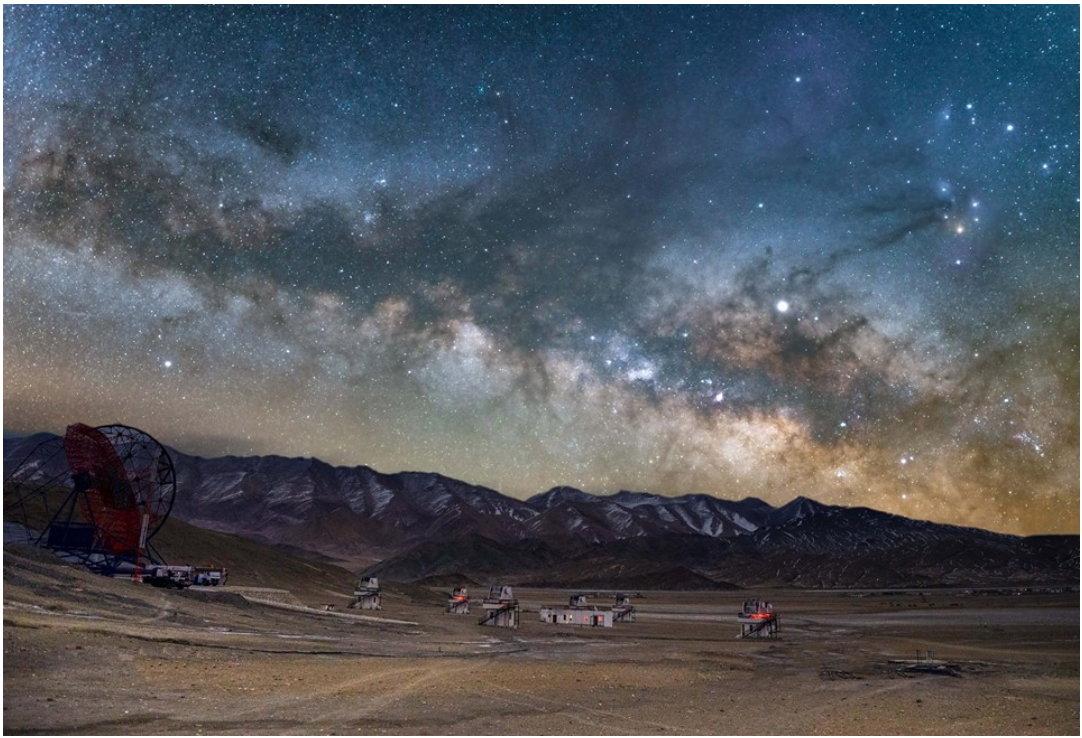

Figure 1. (e) The 2 m optical/NIR telescope (IIA), in operation since 2003. (f) The 4300 m high base area where the 7-element HAGAR TeV Cherenkov array is operational since 2008 and the large 21 m MACE Cherenkov telescope is under commissioning. (image credits: Dorje Angchuk).

The Ladakh region lies north of the Great Himalayan and Zanskar ranges, which greatly moderate the effects of the Indian monsoons. Annual precipitation at Hanle is < 70mm (50mm of rain and 15cm of snow; Prabhu, 2008; Cowsik, Srinivasan & Prabhu, 2002), lower than Mauna Kea and similar to Chajnantor. The median wind speed is 3 m/s (night: 2 m/s), substantially lower than both Mauna Kea and Chajnantor, with a maximum of 32 m/s (Bhatt, Prabhu & Anupama, 2000). Site testing studies using 220 GHz atmospheric opacity measurements over a three year period, jointly carried out by the Raman Research Institute, the University of Tokyo and the Indian Institute of Astrophysics, show that the site is comparable to Mauna Kea overall, with better  winter months (Ananthasubramanian et al. 2003, 2004). The annual median 220 GHz opacity is 0.11, substantially better at 0.09, with the Jul/Aug summer months excluded. During the Oct-Mar winter months, the 220 GHz zenith opacity quartiles are: 0.05, 0.07 and 0.1.  For comparison, the quartiles of the opacity at 225 GHz over the best six months are 0.05, 0.08 and 0.14 for Mauna Kea and 0.03, 0.04 and 0.07 for Chajnantor (5000 m; Chamberlin et al. 1997; the 225 GHz opacity is expected to be ~ 3% worse than at 220 GHz for Chajnantor; Radford et al. 2001). The dry periods show little diurnal variation, thus both the night and day times are usable for sub-mm observations, easing seasonal constraints on target observability. A study using the MERRA-2 reanalysis data covering longer and current periods shows similar transmission characteristics (Raymond et al. 2019). Previous scouting visits by the staff of the CfA Receiver Laboratory identified the 5000 m Pologongka La location as a potentially good higher site. The last 10 km stretch of the road to the Pologongka La site is currently unpaved.  A planned new rail road line passes within ~ 20 miles of this site and includes a station at a distance of 25 miles (40 km) by road.

## 3. CfA Sub-mm/THz Efforts

A brief outline of sub-mm/THz efforts at the CfA is provided here, as a detailed discussion is not necessary. The Smithsonian Astrophysical Observatory (SAO), a member of the Center for Astrophysics | Harvard and Smithsonian (CfA) is a pioneer in mm/sub-mm/THz astronomy and technology. The staff of the SAO and the CfA (1) developed, deployed and currently operate the first sub-mm wave interferometric aperture synthesis telescope (the Submillimeter Array, SMA), on Mauna Kea, Hawaii (Ho et al. 2004; Blundell, 2007). The SMA commands high oversubscription rates for observing time (~ 3, up to ~ 7 including large scale projects; Sridharan, 2018), its capabilities having been continuously upgraded, with the wide band SMA (wSMA) expansion for the receivers and correlator currently under way (2) made the first ground-based THz astronomical observations using a telescope developed and deployed by SAO's Submillimeter Receiver Laboratory at 5,525 m altitude in northern Chile (Blundell et al. 2002; Marrone et al. 2005) (3) carried out the largest surveys of molecular clouds, mapping the entire Milky Way Galaxy (the CfA CO survey; Dame et al. 2001), with on-going limited expansion to other species (4) led the formation of, and studies with, the Event Horizon Telescope (the EHT) which imaged for the first time, the signatures of the Super Massive Black Hole at the center of M87, an achievement for which the worldwide EHT collaboration was awarded the 2020 Breakthrough Prize in Fundamental Physics (The EHT Collaboration, 2019a,b,c,d,e,f). These efforts were enabled by broad expertise in mm/sub-mm wavelength instrumentation and astronomy in general and in the area of high sensitivity superconducting (SIS) front end receiver technology, a field in which the SAO sub-mm wave receiver laboratory is a leader.

## 4. ISRO-CfA Collaboration proposal

Currently, ISRO is exploring high sensitivity SIS receiver technology development under its previously mentioned sub-mm/THz initiative (section 1). As part of this endeavor, SAC-ISRO has sought participation from the SAO Receiver Laboratory, through collaboration. In return, opportunities to jointly shape and pursue new science from the high Himalayan sites are offered at little cost. As the overarching goal of the CfA is science, we evaluated the science prospects of the opportunity presented by this collaboration proposal. We identified two broad areas of interest to the CfA community, with significant current and future possibilities, discussed in more detail below. Given the science opportunities of interest to the CfA, created by SAC's plans to construct and operate a sub-mm wavelength astronomy observatory in the Himalayas, and SAO's extensive expertise in developing sophisticated instrumentation for astronomy, the staff of the SAC, RRI, and SMA/SAO engaged in informal discussions on possible collaboration. These discussions led to mutual informational visits and wider discussions with other Indian institutions, eventually resulting in the formation of the Indian Sub-millimetrewave Astronomy Alliance, ISAA, to pursue the sub-mm initiative on a national basis in India. The ISAA partnership brings together scientists and engineers from SAC, ISRO, Raman Research Institute, Tata Institute of Fundamental Research, Indian Institute of Space Science and Technology-ISRO and the Indian Institute of Astrophysics. Further development of the submm/THz initiative is proposed to be jointly pursued between the CfA and ISAA, with SAC playing a lead role in observatory and instrumentation development. The other ISAA institutions will provide science support and contribute relevant technical expertise they possess. The IIA will also provide logistical and site infrastructure support. As described before, ISRO has a track record of working with national and international academic and research institutions, and the sub-mm initiative is being positioned as a national pursuit. In the following sections, we discuss the two broad science areas identified from the CfA perspective, in more detail. These are areas where significant interest currently exists at the CfA.

## 5. Science Opportunities

## 5.1. The *CO-Dark-H2* phase of the ISM

The central process in galaxy evolution is the life cycle of the ISM – the warm ionized medium is thought to condense into diffuse atomic gas, which in turn into the dense molecular ISM, followed by clump, core and star-formation. The stars inject energy and momentum into the natal ISM, leading to the destruction and dispersal of the dense phase through radiation, stellar winds and supernovae, back into the diffuse medium. To fully understand the operational processes, it is necessary to observe all the ISM phases involved. The diffuse neutral medium is directly observed through the 21-cm fine structure spin transition of atomic hydrogen, HI. The dominant constituent of the cold dense medium, the active star-forming component, $H_2$, is not directly observable and is usually traced by its easily observed proxy, CO, the second most abundant molecule. The lowest rotational transition of CO is in the 3-mm atmospheric transmission window and has been extensively mapped, the CfA CO survey being the largest and most complete, covering the entire Milky Way. CO is also the most widely observed extragalactic molecular tracer. These surveys form the basis of our understanding of molecular clouds and their characteristics and distribution in the Milky Way and external galaxies. However, estimates of the amount of $H_2$ from CO maps consistently fall short of estimates from other tracers - the Galactic diffuse gamma ray emission (Gernier, Casandjian & Terrier, 2005; Wolfire, Hollenbach & McKee, 2010; Ackerman et al. 2012; Remy et al. 2017, 2018; ) and continuum dust emission (Planck Collaboration, 2011; Remy et al. 2018), which implies that our understanding may be incomplete. As explained below, this mismatch may be at the center of the transition between the diffuse and the dense phases – the formation of the molecular clouds - a process that has not been adequately observed or understood. Similarly, a component of the neutral atomic phase not traced by 21-cm HI observations may also be present (Planck Collaboration, 2011; more below). We note that less extensive observations of other species sensitive to higher densities and a range of physical and chemical conditions towards targeted regions have allowed explorations of the rich fields of star-forming cores, the star and planet formation and their attendant processes, and astrochemistry.

The proxy function of CO has a number of limitations, outlined below, leading to the aforementioned discrepancy. (1) As CO lines are optically thick, the bulk of the CO gas is not directly observed and $H_2$ mass estimates use the empirical and statistical CO to $H_2$ conversion factor $X_{CO}$, appealing to virial equilibrium, generally valid for $M > 10^5 M_\odot$ and normal CO abundance (e.g.: Bolatto, Wolfire & Leroy, 2013; Papadopoulos, Bisbas and Zhang, 2018). This conversion, being statistical, is not necessarily applicable to every individual cloud. (2) Far-ultraviolet (FUV) illumination of molecular clouds leads to a CO deficit due to photodissociation whereas $H_2$ is self-shielding. The extent of FUV penetration depends on metallicity which determines the amount of dust, the main absorber. Turbulence and clumpiness in molecular clouds allow FUV to reach beyond the surface regions (3) Cosmic rays (CR) penetrate deeper unobstructed, and destroy CO, a role recently realized to be central in regulating the CO-to-$H_2$ ratio (Bisbas et al. 2017; Papadopoulos et al. 2018). CO dissociation produces atomic carbon (C) in these regions. Together, the CR and FUV destruction of CO is prevalent throughout molecular clouds, extending to the surfaces of, and into, the clumps and cores. This is consistent with the absence of stratification in available limited atomic carbon line observations and their general similarity to CO observations, contrary to initial expectations. (4) The CO formation rate is slower than that of $H_2$ (Suzuki, 1992; Bergin et al 2004), leading to regions with insufficient CO to trace $H_2$, especially in early stages of molecular cloud formation. In summary, significant amounts of $H_2$ may exist without corresponding CO, a phase termed the *CO-dark-$H_2$*. The undetected $H_2$ component may be as large as that accounted for by observed CO (Grenier, Casandjian & Terrier, 2005; Planck Collaboration, 2011), may be widely variable, and is poorly characterized.

Theory, simulations and limited observations identify atomic carbon C to be a good tracer of this hidden phase and indicate differences with molecular gas distribution derived from CO maps. Specifically, neutral atomic carbon CI traces an additional more diffuse component compared to CO (Fig. 2). Dissociation of CO by FUV produces atomic carbon in its neutral and ionized forms CI and CII, while CR induced dissociation largely results in CI (for T < 50 K; Papadopoulos, Bisbas and Zhang, 2015). There are two fine structure lines of CI, $^3P_1 - ^3P_0$ and $^3P_2 - ^3P_1$, at 492.2 and 809.3 GHz, respectively. With upper energy levels of 24.2 K and 62.5 K, and critical densities of 500 and 2500 cm$^{-3}$, they are easily excited, more so the 1-0 line, typically stronger than the 2-1 line by a factor of $\sim$ 2. The CI lines are optically thin in most circumstances. Therefore, all of the emitting gas is observed and the mass depends linearly on luminosity, which is a clear advantage, compared to CO. Also, unlike the CO to $H_2$ conversion factor, $X_{CO,}$ the CI to $H_2$ conversion factor, $X_{CI,}$ is not a strong function of time in metal poor systems (Glover & Clark, 2016). Thus, CI is an attractive tracer from multiple perspectives. The lowest line of ionized carbon CII is at 158 micron (1.9 THz) and is not easily accessed from ground. The NASA GUSTO balloon borne survey includes this line and is set to begin in Dec 2021.

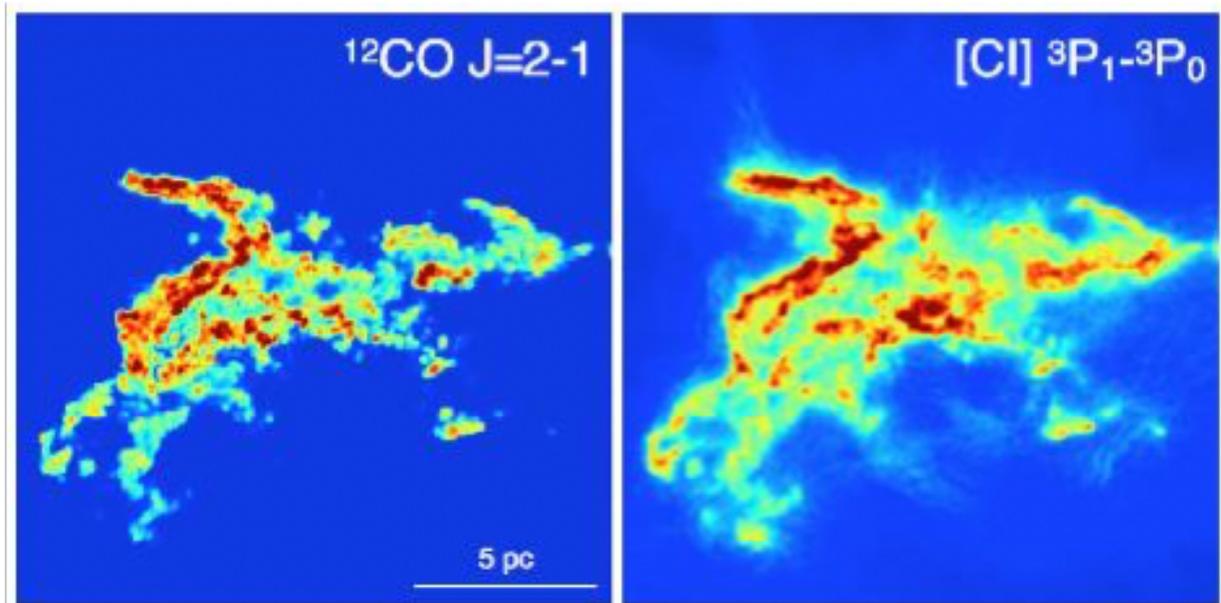

Figure 2. As can be seen in the simulated CO (2-1) and CI (1-0) integrated intensity images, CI traces an additional, more diffuse component of $H_2$ that is missed by the CO observations (Glover et al. 2015). While the maps are similar, the material missed by the CO observations is in interface regions with the diffuse ISM where the cloud and clump formation processes are active (from simulations by Molina and Glover; image from Stacey et al. 2018).

Observations of these CI lines require very high, dry sites, such as in the Himalayas, to overcome atmospheric absorption, which has limited widespread studies. A primary focus of the submillimeter initiative proposed here will be the 492 GHz CI (1-0) line, as it is brighter in addition to the transmission being better than at 809 GHz where the (2-1) line lies. To this end, it is proposed to develop a new 492 GHz receiver that can also simultaneously observe the CO (4-3) line at 461 GHz in its lower side band (Figure 5). This CO line would allow the characterization of excitation conditions in the observed regions (section 5.3).

We note again that cold atomic hydrogen that is not traced by the 21-cm HI observations due to high optical depth, but that is also not dense enough to form $H_2$ or CO, may exist and could be traced by CI emission (Burton et al. 2015; Fukui et al. 2015). The Planck Collaboration (2011) found evidence for such an ISM component by comparing 21-cm HI data and $350\mu m - 2mm$ continuum dust emission data. This is a largely unexplored area and CI observations may be the only way to study such regions in the Galactic plane, and their kinematics.

In the next sections, we address the importance of observing *CO-dark-$H_2$*, through CI observations. The surface regions of molecular clouds and clumps and their interfaces with young stars are dominated by *CO-dark $H_2$* and are missed by CO observations. CI studies will have significant impact in several areas, Galactic and extragalactic, as discussed below.

### 5.1.1. Galactic Science

*1. Molecular cloud formation:* The processes that form the molecular clouds – the cold, dense, star forming medium – are unclear. In addition to other paths, this may happen through (a) the gradual gravitational collapse of the diffuse atomic medium (b) the intersection of expanding sheets/shells of diffuse medium due to energy and momentum input from a previous generation of stars (e.g.: Dobbs et al. 2014). These processes, responsible for the transition of the atomic medium into a molecular phase, and the processes that lead to the formation of clumps and cores in them, operate predominantly in the outer regions of these forming molecular gas structures and are thus not fully accessible to current molecular line tracer observations, like CO. This is because, these regions lack adequate amounts of the tracer species, as described before (section 5.1). In addition to observing the missing material, by up to a factor of ~ two, the density, temperature and kinematic (flows) structures in these regions, accessible to CI observations, are critical to understanding their formation mechanisms which cannot be explored by other means. The kinematic and mass information in these observations can provide a path to quantify the growth rate and its sign – aggregation and destruction – of molecular clouds.

*2. Scaling Relations:* CI observations can inform scaling relations in molecular clouds – e.g., star formation rate versus mass (e.g., Lada et al. 2012) - by tracing new material. The missing material can contribute to the scatter in the scaling relations for molecular clouds and structures within them. Understanding this scatter is a current focus and important to better isolate the processes central to cloud, clump, core and star formation. The extragalactic scaling relation, typically studied with CO data due to insufficient observations of higher density tracers, suffers from unknown and varying amount of missing material. The characteristics of scaling relations using CI observations is an unexplored area and needs to be studied both in the Milky Way and in external galaxies. This can be extended to high-z objects as more redshifted CI data become available. As CO suffers for the reasons mentioned before and higher density tracers are too weak to detect, CI observations may be the only available approach to extend the scaling relations to high redshifts. This makes CI observations in the Milky Way and in nearby galaxies a critical need.

*3. Diffuse Gamma Ray Emission:* The CI observations can provide an additional new template in decomposing the diffuse gamma ray emission into its components. The Galactic gamma ray emission is not yet understood in full detail in terms of the known processes which create it. After subtracting spatial templates for the known processes, the residuals are sought to be understood through new processes (e.g., Balaji et al. 2018; early work to isolate extragalactic emission: Sreekumar et al. 1998) - the Fermi bubbles (Dobler et al. 2010; Su, Slatyer & Finkbeiner, 2010; Finkbeiner, Su & Malyshev, 2014) and potential dark matter annihilation/decay (Goodenough & Hooper, 2009; Daylan et al. 2016) being examples. It is necessary to improve the templates that account for the known material to better isolate and constrain the unknown new processes. This study will require the use of the best weather time (see section 8).

*4. Star-forming regions:* In the regions immediately surrounding newly formed stars, UV radiation is strong, leading to photon dominated regions (PDRs). Atomic carbon is a good tracer of gas in these regions, in understanding the processes at the interfaces of stellar winds and the natal material at the walls of the outflow cavities. For the modest size of the antenna being considered, we will be limited to high-mass star-forming regions, as the lines towards low mass regions are weaker.

*5. Astrochemistry:* Atomic carbon is an intermediate species in the astrochemical networks that form carbon bearing molecules in these interface regions. A significant part of astrochemistry occurs in the PDR regions with different time scales, products and dominant pathways compared to denser, more UV shielded environments. Among the variety of astrophysical contexts where PDRs appear, massive star-forming regions, being bright, will be readily accessible to this initiative. CI line observations of such regions can lead to understanding of the chemistry and physical properties of the regions. The direct observation of atomic carbon can inform and constrain the chemical models. Limited observations show that the C/CO ratio varies by a factor of ~ 4 over the Milky Way molecular clouds and high mass star forming regions, with a possible dependence on evolutionary stage (Buether et al. 2014). The C/O elemental ratio and C/C$^+$ ratio play important roles, with C/O influencing the production of complex organics such as fundamental pre-biological complex nitriles (Le Gal et al. 2019), which in turn are thought to have played a key role in the emergence of life on Earth (Patel et al. 2015). Investigations in this area are just beginning, and clearly, more observations covering larger samples and a range of conditions are warranted.

As previously stated, since the C emission depends on metallicity and CR and FUV environments, to adequately address the topics outlined above, it is necessary to make large scale maps of the Galaxy and of individual clouds, covering a range of Galactocentric radii and conditions. A small dedicated telescope is well suited to such a task.

### 5.1.2. Extra-galactic Science

*1. Local galaxies:* Fine structure lines of abundant metals, like CI, CII, OI, NII and high J rotational transitions of CO dominate the line cooling of molecular clouds. Of these, CI and CO (4-3) will be accessible to the proposed instrument. CI is the brightest line emitted by molecular clouds and is thus an important line probe for high-z galaxies, redshifted into mm- and cm- wave bands, increasingly becoming an important tracer with the SMA, NOEMA and ALMA interferometers. The mapping of these lines in local galaxies and in the Milky Way, at linear resolutions comparable to extragalactic interferometric observations is necessary to better understand, and as bench marks for, the high-z observations. Therefore, large scale maps of the local galaxies covering their full extent with single dishes are valuable and possible. Thus, observations of our nearest neighboring spiral galaxies M31, M33 which harbors a large metallicity gradient – ranging from ~ solar in the central regions to 1/10 at the edges (Vilchez, 1988), and M51, are important goals. These nearest galaxies happen to be northern, M51, M31 and M33 at declinations +47°, +41° and +31° respectively, ideally suited to the Ladakh sites at ~ 33° N latitude. At the distances of M31 & M33, the resolution provided by the proposed 3 – 6 m antenna approaches GMC scales (100 - 200 pc) and is limited to intra-galaxy 1-2 kpc scales in M51, similar to scales accessed by interferometers in more distant objects. Observations of M31 may also offer synergies with the on-going SMA Large Scale Project to study dust continuum emission from individual extragalactic molecular clouds for the first time (Lada & Forbrich et al. 2018), especially if the SMA is equipped in future for this band (section 6). A 6 m antenna is preferable for these observations. As stated before, due to the dependence of the C emission on metallicity, and CR and FUV environments, it is necessary to make large scale maps of the full extent of the nearby galaxies, the Milky Way and of individual clouds, covering a range of Galactocentric radii and conditions. Maps of individual clouds in the Milky Way would inform interferometric maps of local galaxies, with similar linear resolution.

*2. Dwarf Galaxies:* The low-metallicity ISM of star-forming dwarf galaxies has very different properties than the ISM in massive disk galaxies like the Milky Way. Due to the reduction of dust-shielding and self-shielding, photodissociation regions (PDRs) are expected to be much larger in the metal-poor ISM, while CO emission is faint. In the N159 and 30 Doradus HII regions in the LMC, this expectation is confirmed, as the [CII]/CO and [CI]/CO luminosity ratios are enhanced compared to star-forming regions in the Milky Way (e.g., Israel et al. 1996; Stark et al. 1997). However, it is unknown whether these regions are representative of the entire metal-poor LMC, and a systematic survey of CI in dwarf galaxies has yet to be conducted. In addition to understanding PDRs in low-metallicity environments, another critical question concerns the amount and distribution of the CO-dark-H2 phase in dwarf galaxies. Yet only a handful of CI observations of dwarf galaxies exist (e.g., Stark et al. 1997; Gerin & Phillips 2000; Pineda et al. 2017), and no Local Group dwarf galaxy has been fully mapped in CI. With an apparent size in the visible of 7′x 6′, IC 10 (declination +59°) would be the ideal target to fully map from the Ladakh site. This metal-poor dwarf ($Z \sim 1/5\ Z_{sun}$) has been previously mapped in CO (Leroy et al. 1996), but not in CI. Other northern targets that could either be fully mapped or observed with single pointings include NGC 1569, Leo A, Mrk 209, NGC 2366, and NGC 4214.

3. *Ultra-luminous Infrared Galaxies (ULIRGs):* Systematic extragalactic CI observations are very limited, to about a few tens of targets. The most extensive published survey observed a sample of 76 ULIRGs with Herschel-SPIRE FTS for CI (Israel et al, 2015) at low spectral resolution (1.2 GHz, 720 km/s), leading to spectrally and spatially unresolved detections. Due to the inadequacy of CO, given the strong UV radiation, CI observations are relevant in these galaxies. Based on the Herschel study, which concluded that neither CO nor CI is a good predictor H2 masses, these galaxies are bright enough for a 3-6 m class antenna. As a majority of these observations will be spatially unresolved detections, a 6 m antenna is preferred to maximize collected flux. Understanding the utility of CI as a tracer of H2 in the extragalactic context, and the study of scaling relations using the CI data as previously mentioned, are areas in their beginning stages with little previous work. Clearly, studies of larger samples at better spectral resolution are needed. The CO(4-3) line observed simultaneously along with lower J transitions will allow kinematic comparison of CI and CO, precluded in the unresolved Herschel study. These observations will also be pathfinders for high resolution observations with ALMA, and with the SMA in future if equipped for this band (section 6). Due to the presence of multiple atmospheric absorption bands immediately below the 492.2 GHz line (Figure 5), the distances covered in such studies will have gaps as the CI line is redshifted into these bands, which cannot be overcome by ground based observations.

We note that lines of CII, NII, OI and higher J CO will be accessible from higher sites available in the Ladakh region, although not in the bands proposed in the immediate goals.

### 5.1.3 A Sampling of Existing Observations and Summary

The characteristics of CI emission, as seen in the limited observations available, point to missing material, but there is wide variation and a lack of consistency. The most recent observations and modeling of the CI (2-1) observations from the High Elevation Antarctic Telescope (HEAT) suggest significant amount of CO-dark-$H_2$. A 30% missed material is estimated for the peripheries of G328 with a total missing mass of 67% for the full volume (Burton et al. 2015). For G332, the number is 16% for the peripheries which would translate to $\sim$ 30% for the full volume (Romano et al 2019). The distribution of estimated missing mass fraction for G332 for assumed excitation temperatures, shown in Fig. 3, is seen to be in general agreement with the simulation in Fig 2. Higher resolution observations with the 10 m ASTE telescope found a few clouds in the Central Molecular Zone to have a large CI/$^{13}$CO ratio, pointing to large missing masses (Tanaka et al. 2017), with possible youth of the clouds and CO destruction by X-ray and CR offered as potential causes. Older AST/RO observations of the Galactic center region found CI with similar spatial extent as low-J CO but more diffuse and specific ratio estimates were not made (Martin et al, 2004). Early CSO observations (2.2′ beam; Jaffe et al. 1996) found an increased CI/$^{13}$CO in the inner 6 pc. In Orion A and B, using 2.2′ resolution observations with the Mt. Fuji 1.2-m telescope, Ikeda et al (2002) found that the CI/$^{13}$CO column density ratio varied by a large factor of 0.2 to 2.9 over the cloud edges, while generally similar in the central regions with a factor of 2 variation. The CI/$^{13}$CO column density ratio and the integrated intensity ratios are both higher towards the edges by a factor of $\sim$ 2-3 compared to the central regions. Limited ASTE observations of Orion A at higher resolution (20″) show similar results (Shimajiri et al. 2013). In closing, as outlined in the above incomplete summary, there is an important need to observe large areas with uniform observing parameters. Thus, a CI survey of a full outer quadrant of the Galaxy and selected molecular clouds can allow important advances. Observations of the outer Galaxy are critical for including the effects of metallicity variation with Galactocentric radius.

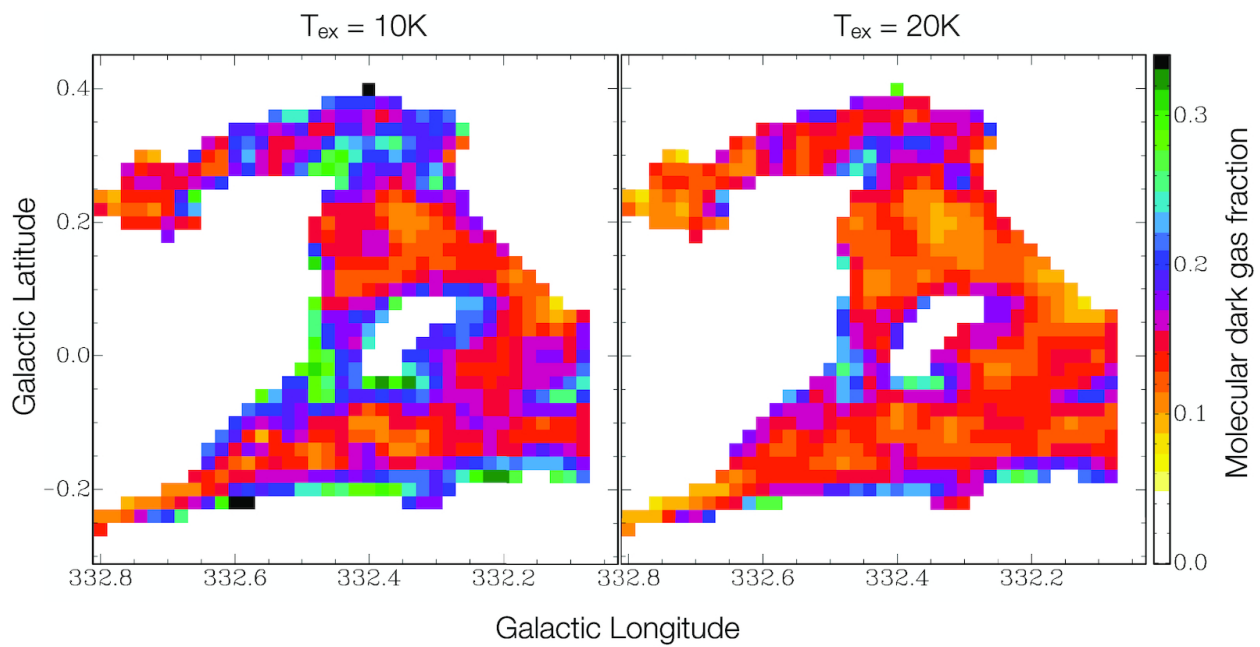

**Figure 3.** *Dark molecular gas* fraction map across the molecular ring in G332, derived from HEAT observations. A fixed Tex = 30 K was used for the [C I] column density while for CO two Tex values were used, as indicated above each panel. The scale is the same for both pictures (from Romano et al. 2019).

The CO-dark-$H_2$ part of the science case has a lot in common with those of the 6 m Cerro Chajnantor Atacama Telescope – prime (CCATp) telescope (Stacey et al. 2018), and the 0.6 m High Elevation Antarctic Terahertz telescope (HEAT; Kulesa et al. 2011). While not as fast or sensitive as the CCATp, the modest Himalayan antenna is intended to carry out dedicated CI (1-0) surveys. The $2^{nd}$ quadrant of the Milky Way transits at < 30° elevation from the CCATp latitude, with the outer Galaxy longitude range ~ 90° − 160° transiting below 20°. The Ladakh latitude provides complementary coverage (see section 7 below). The proposed Himalayan antenna will deliver better spatial resolution than the 60 cm HEAT. The HEAT telescope recently ceased operations and was removed in Jan 2019. Results from completed CI (2-1) HEAT observations, covering limited strips of the Galactic plane are beginning to appear. The emerging understanding so far is that there are pervasive, diffuse molecular clouds not seen in existing CO and HI maps (Papitashvili, 2019). This clearly points to the need for concerted efforts to carry out large scale CI surveys. More importantly, in the northern hemisphere, the Himalayas are the only place from where these studies can be carried out efficiently, apart from the Greenland ice sheet, and thus the Himalayan sub-mm telescope holds unique potential. The $2^{nd}$ outer Galaxy quadrant, with no planned surveys by other telescopes, and the nearest northern spiral galaxies are niche areas. The large-scale CO (4-3) survey carried out simultaneously in the lower sideband along with maps of lower J CO transitions can constrain excitation conditions over the same large areas covered. We note that although SOFIA has receivers for this band, the CI line is not a science focus as it is also accessible from ground (Menten, K; Gusten, R, private communications). In addition,  as SOFIA observing time is very limited, large surveys are not possible.

## 5.2. Event Horizon Telescope/mm & sub-mm VLBI

The recently achieved first imaging of the signatures of the super massive black hole at the center of M87 is a landmark CfA-led development and has opened up an observational approach to studying fundamental phenomena near black holes (The EHT Collaboration, 2019). The award of the 2020 Breakthrough Prize in fundamental physics to the worldwide EHT collaboration for this achievement is a recognition of its seminal importance.

The highest resolution possible for earth sized baselines at 230 GHz has been reached with the current EHT configuration. In its next generation, the EHT now targets improvements to imaging quality, higher frequency observations and dynamic imaging as its next steps (Doeleman et al., 2019; Blackburn et al. 2019). Given the sparseness of the EHT station distribution, the *uv* coverage is limited, which impacts imaging quality.  To address this, EHT expansion to additional stations is being explored.  A VLBI station in the Himalayas, a longitude range lacking current EHT coverage, enables advances on this front and will also be able to support the second goal of higher frequency operations. Such a station will have mutual target visibility with some existing stations - NOEMA, IRAM 30M (PV), GLT, SPT and the SMA, with a very small overlap at low elevations with ALMA - for some of the key targets - e.g.: M87, OJ287, and Sgr A*. The best EHT enhancement is expected for M87 due to its favorable declination. Figures 4(a) and (b) show the visibility overlap for M87 with other current
stations and the *uv* coverage improvement. By definition, the pursuit of the highest resolution possible with VLBI means that the longest east-west baselines are able to observe a target only for a short period of time and at low elevation angles, and not all baselines will be able to observe all targets. Thus, an expansion of the station distribution brings significant benefits. Even without adding unique spatial frequency coverage, additional stations bring (1) weather resilience, increasing the likelihood of successful observations, which must be planned well in advance with limited predictability of joint weather conditions across the diverse EHT stations (2) more complete continuous interferometric visibility coverage with time (better snapshot coverage, in other words), improving the fidelity of short integration images, important for successful dynamic imaging (3) expand the geographical footprint, providing greater time overlaps with other observatories for multiwavelength variability observations (4) provide the most easterly EHT station - the SMA-Hanle baseline would be the first to acquire targets in a typical VLBI observing run and (5) provide calibration redundancies. In the context of other additional stations being pursued, a Himalayan station enhances the EHT with additional baselines and closure triangles.

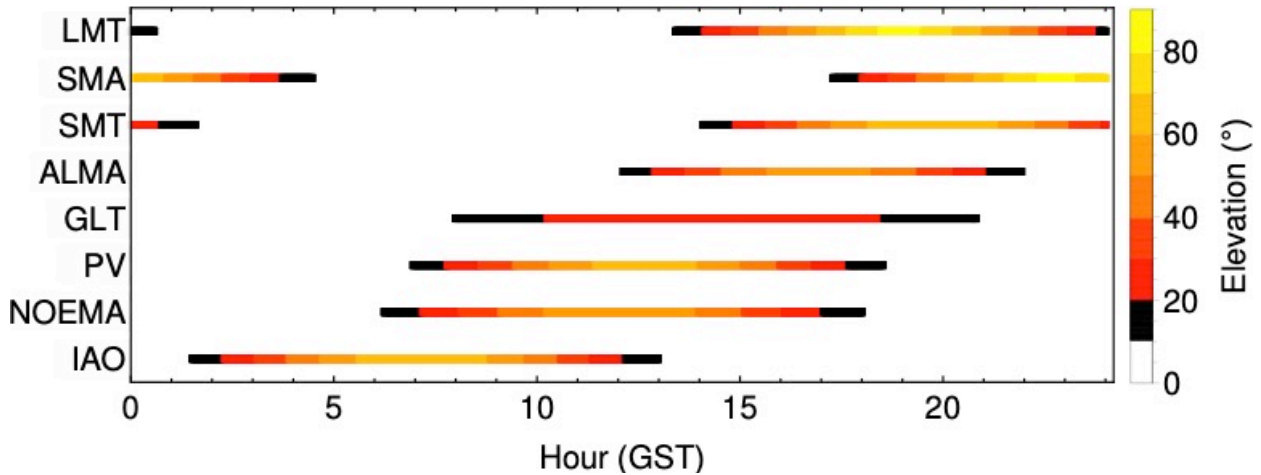

Figure 4. (a) Target visibility overlap on M87 for various current EHT stations with the location of the proposed sub-mm antenna at the Hanle, Indian Astronomical Observatory (IAO) site.

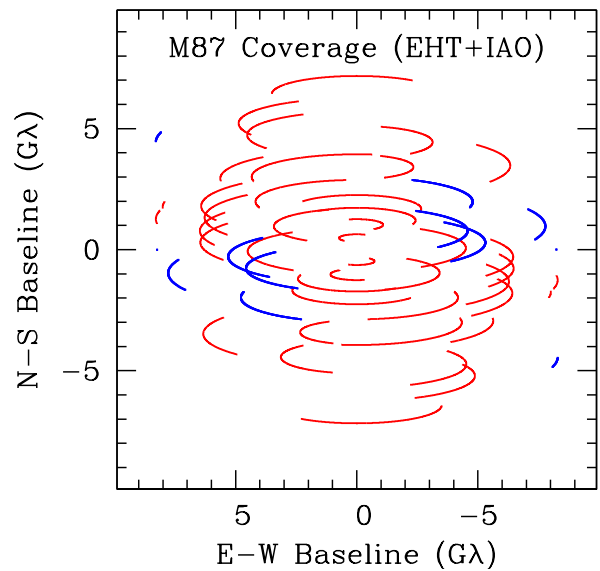

Figure 4. (b) Interferometric visibility coverage obtained through earth rotation for the current EHT configuration on M87, with the IAO addition shown in blue.

We estimate that the noise level for the most sensitive baseline for M87 observations, viz., the phased NOEMA-IAO, will be 20 mJy for a 10 s integration for a 6 m antenna (40 mJy for a 3 m antenna) with the expected flux being 80 mJy. A 5 $\sigma$ detection can be achieved in 15 seconds. With the SMA, a 1 $\sigma$ of $\sim$ 50 mJy is expected for 10 seconds and a 5 $\sigma$ detection can be obtained in $\sim$ 2 minutes. Thus, with the expected days timescale variability in M87, a Himalayan station contributes to the dynamic imaging capability in addition to improving image quality. Sgr A* is less well located and the impact will not be significant, but every enhancement contributes to imaging improvements. It is necessary to have detections on time scales short compared to atmospheric coherence times for successful VLBI. Therefore, a 6 m antenna is clearly preferred over a 3 m antenna. A larger 10-12 m class antenna will make the EHT prospects very attractive. A larger antenna would be particularly important for 345 GHz observations, a strength of the Himalayan sites, where the sensitivities are lower and stability time scales are shorter.

The proposed EHT-VLBI work would involve equipping the observatory with 230 & 345 GHz receivers and even higher frequencies in future. The station will also have to be provisioned with high speed recording equipment and a maser reference. In the longer term, considering ISRO's core competency being space capabilities, combined with interest and increasing emphasis on space sciences, partnering in a future space VLBI experiment is a possibility. A space based VLBI station can fill in the large gaps in the sparse *uv* coverage provided by an earth based VLBI array and do so quickly, as it is not locked to earth rotation speed, unlike in earth rotation aperture synthesis (Palumbo et al. 2019; Johnson et al. 2019). This will allow dynamic imaging of short term variability, such as expected in the Sgr A* SMBH. Initial pursuits in the Himalayas, with this long-term goal in mind are well worth the effort and resonate with nascent CfA initiatives. In this context, we note that ISRO has recently equipped the upper stage (PS4) of its workhorse launcher, the Polar Satellite Launch Vehicle (PSLV), for orbit keeping, maneuverability, power and telemetry functions on an experimental basis. This makes the spent PS4 stage available as an orbital platform for short experiments, after the primary payloads have been delivered. We are in encouraging preliminary exploratory discussions with the Space Infrastructure Programme Office (SIPO) of ISRO to utilize this platform to carry out possible experiments to assess and develop technology readiness level (TRL) of subsystems, towards an eventual space VLBI station – e.g., proving the performance of reference sources. Potential laser communication experiments are also being contemplated, to enable high data rates both for space VLBI and, in he near term, for the south pole EHT station. The polar orbit of the platforms is particularly well suited for south pole data transfer. With several launches a year, multiple platforms may be available simultaneously, allowing more experiments for flight testing other subsystems including simple potential connected element experiments.

Equipping the Himalayan antenna for the 3-mm band will enable participation in the Global Millimeter VLBI Array (GMVA) observations. Due to the availability of Asian stations for this band, a Himalayan station will add additional baselines, *uv* and time coverage, while also providing baselines with more westerly stations in Europe and the GLT. If pursued, this will enhance  and solidify CfA's GMVA presence.

## 5.3. Nominal Weather Science

The 461/492 GHz observations require very good weather conditions for observations, as the lines are close to an atmospheric transmission dip, due to the 487 GHz $O_2$ line (Figure 5). The lower frequency receivers in the 230 & 345 GHz bands needed for EHT science, with observing runs taking up limited stretches of time,   can be used to carry out large scale surveys in the CO (2-1), (3-2) and other molecular lines in the Milky Way and nearby galaxies such as M31, M33 and M51. These observations, combined with the 461 GHz CO (4-3) data from the lower sideband and (1-0) data from the CfA CO survey should be able to provide a comprehensive assessment of excitation conditions in the regions covered. This would allow characterization of molecular excitation  over unprecedentedly large areas with a range of conditions. Spectral line surveys of selected targets are also possible. The Indian side has expressed particular interest in the Milky Way molecular line mapping and spectral line surveys in these bands. It is expected that about half the time in the. winter months (Oct. – Mar.) and a majority of the time in the rest of the year will be available for these observations.

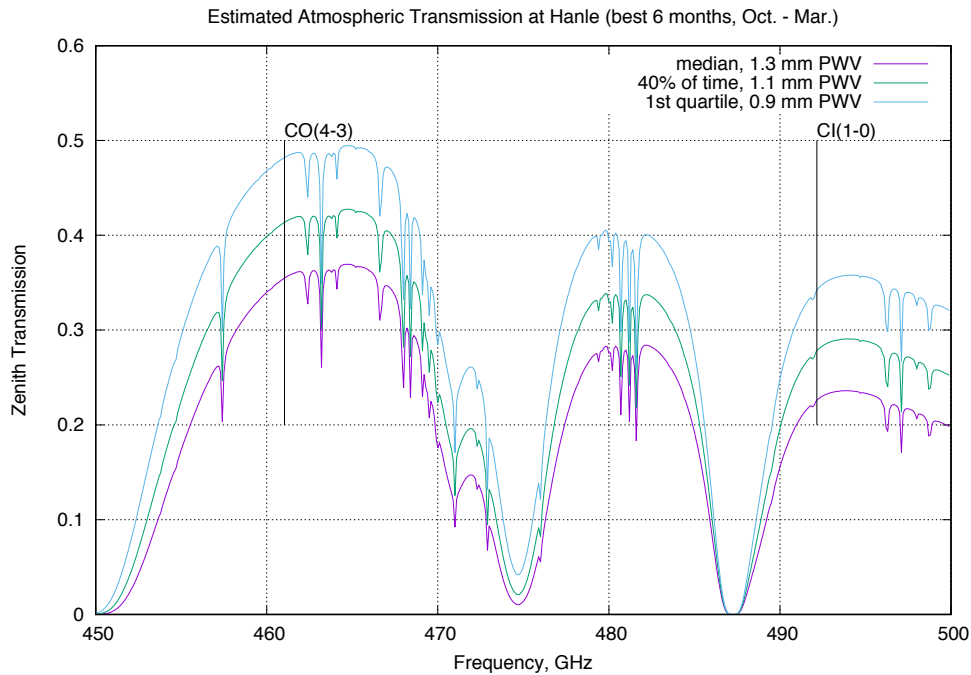

Figure 5. Atmospheric transmission for Hanle (4500 m), estimated from measured 220 GHz opacities and relations and models applicable to Mauna Kea (4000 m; Dumpsey et al. 2013).

## 5.4. Additional Science

The key science topics outlined above are not exhaustive and opportunities for a number of other investigations exist. We outline a few potential avenues below.

Episodic accretion has increasingly become a mainstream idea as an explanation for the protostellar luminosity problem. Young stars are found to be less luminous than would be implied by the average accretion rates required to build them. Periods of high accretion would alleviate this problem (Kenyon et al. 1990; Hartmann, Herczeg & Calvet 2016). Recently, in the case of high mass stars, there have been observations of dramatic increases in luminosity and appearance of new maser lines (Brogan et al. 2019). Monitoring observations of a set of lines at frequent intervals is necessary to catch these events to enable high resolution studies of their earliest phases. The proposed telescope is well suited to monitoring methanol maser lines in the sub-mm bands as part of a global network of observatories participating in such observations.

Spectral line surveys carried out towards a limited number objects have brought a wealth of information on astrochemical processes. The field still lacks systematic studies of samples of objects in different classes to study the variation of chemical characteristics. This is another area where the Himalayan sub-mm telescope can contribute by extending such surveys to statistically viable samples.

Finally, this telescope can also be used to enlarge mapping surveys to species other than CO and its isotopologues, in addition to the CI survey proposed.

Turning to earth science, there is significant capability at the SAO receiver laboratory in modeling atmospheric lines (Paine, 2019). Specifically, modeling support to derive atmospheric moisture and temperature profiles using data from the water vapour sounder payload which the SAC group has developed, expected to fly in the near future, will be a valuable avenue for collaborative work.

## 6. Proposed Instrumentation

To address the goals outlined thus far, the Himalayan submillimeter-wave observatory being developed is proposed to be equipped with high sensitivity, state-of-the-art superconducting (SIS) receivers for 230, 345 and 461/492 GHz operation. Sufficient expertise for providing intermediate frequency (IF) systems and digital backends exists within the ISAA partnership. The details of the final instrument suite are currently under discussion, driven by the science priorities set out in this document. In the current outline, the 230 and 345 GHz receivers will be very similar, if not identical, to the wSMA receivers. They are essential to enable EHT participation and to allow lower J CO and other molecular line mapping and spectral line survey work during times when the atmospheric transmission conditions do not support 461/492 GHz work. The possibility of developing a 2SB receiver upgrade to the wSMA design will be kept in view, which can reduce the required integration times by a factor of up to ~ 4 for spectral line observations and increase signal to noise ratios by a factor of up to $\sqrt{2}$ for the continuum EHT observations. Such 2SB receivers also represent an upgrade path for the SMA. The 492/461 GHz 2SB receiver is a new development targeting the CI (1-0) and CO (4-3) lines in its two sidebands. This receiver will also provide an option for a future new SMA band, enabling high resolution CI and CO(4-3) studies of immediate vicinities of young massive stars and of individual molecular clouds in nearby galaxies. A perusal of the existing limited literature suggests that observations of massive star-forming regions in the Milky Way and brightest parts of nearby galaxies will be feasible with the SMA in the best 10-20% of the weather distribution on Mauna Kea. The presence of the nearest spiral galaxies in the north implies unique access for the SMA. Operation in this new band will also enable prospective future EHT observations at an even higher frequency. A case for the possible addition of this band to the wSMA for high resolution CI observations was discussed in a previous white paper titled *Science with the wideband Submillimeter Array: A strategy for the Decade 2017-2027* (Wilner, 2017).

An antenna of diameter of ~ 3 m is under development and a 6 m diameter is under consideration. While a 3 m size will satisfy the large-scale CI and CO mapping goals, a 6 m size is considered the minimum need for EHT participation and will greatly enhance extragalactic science prospects. A 10-12 m class antenna will be very attractive in this context.

## 7. Collaboration Plan

The basic idea of this initiative and opportunities it represents were presented to the SMA Steering Committee at its May 2019 meeting, eliciting a positive response and a recommendation for further exploration. Accordingly, previous discussions were expanded to include a number of scientists with potential interest at the CfA, both from SAO and the Department of Astronomy, Harvard University, which also met with a positive response and expressions of support and intents of participation. Following up, this white paper has been developed to solidify the ideas, placing them on a more concrete footing for detailed discussions to further develop individual topics. A Memorandum of Understanding between the SAO and SAC-ISRO has been drafted and is undergoing revision cycles between the two sides.

Under this MoU, SAC-ISRO and SAO expect to agree to explore sub-mm wavelength science and engineering projects of mutual interest and benefit. Specifically, SAC and SAO will make the necessary administrative arrangements to enable exchange of technical, engineering, and scientific staff; work together to define the optimum antenna size and first-light instrument suite consistent with science opportunities and budget constraints; jointly explore options for antenna and receiver design and construction; SAO will share its design of the 6 m SMA antennas, as available; and SAC and SAO will work together with the Indian Sub-mm Astronomy Alliance to develop telescope commissioning and science verification plans. Based on progress and continued mutual interest, further agreements are envisaged, regarding the details of the operational phase of the submillimeter wavelength observatory.

Within the broad framework of the MoU, the following detailed plans are being contemplated. A phased approach to fully develop the initiative is proposed to be pursued. In the first phase, scientists and engineers at the CfA and SAC and the ISAA institutions will work jointly to define the science goals and instrument requirements in more detail. In parallel, engineering staff from SAC will spend time at the SAO Receiver Laboratory, participating in the on-going wide-band

SMA (wSMA) receiver upgrade work through the staff exchange program envisaged in the MoU. This will allow the SAC engineers to gain hands-on experience in constructing, testing and operating sub-mm SIS receivers, while contributing technical effort to wSMA build-out and testing. This would provide an efficient approach to collaboration that would benefit both sides in both the short and long terms. Towards the end of this period, one set of 230 & 345 GHz band receivers will be built by the SAC engineers for installation on the Himalayan antenna. Design and development work towards new 461/492 GHz receivers would be jointly pursued during the same time. A 6 m antenna appears to fulfill the science requirements and options for its design construction will be explored. Prospects for a larger 10-12 m antenna will be kept in view. Building a 6 m antenna of the SMA design by Indian vendors is under consideration. If pursued, this would also establish possible sources for spare parts for the SMA antennas. The SAC is experimenting with panel fabrication approaches for a 3 m antenna being developed on a shorter time scale. The second phase targets initial operations, envisaged at the 4500 m Hanle site. While 461/492 GHz observations can be tested and carried out at Hanle, and EHT participation can be accomplished, the full CI mapping science program at full resolution would greatly benefit from a move to a higher site in a later phase. The IIA/IITB experience in the operation of the Indo-US Growth-India 0.7m robotic optical telescope at Hanle can be leveraged in targeting a high time utilization factor for the sub-mm telescope.

We have worked with the SAO export compliance office, arriving at the conclusion that there are no restrictions in effect - India enjoys a preferential status similar to European countries and other allies and is eligible for export with a letter rather than license.

## 8. Sensitivities, Targets and Observing Plans

While both CI (2-1) and (1-0) observations are needed for a proper understanding of CI excitation and for column density derivation, CI column density is quite insensitive to $T_{ex}$, for $T_{ex} > 15$ K (e.g., Ikeda et al, 2002). The proposed survey will initially rely on assumptions, using excitation temperatures from CO observations and existing CI (2-1) observations, where available, for deriving CI column densities. A limited CI (2-1) survey can be carried out in a later phase from a higher site and observations of selected locations can be proposed to SOFIA or other telescopes for better column density determinations and to assess the level of uncertainty in measurements based only on CI (1-0).

At the latitude of Ladakh, limiting observing to an elevation of 40° for the 492 GHz band (winter median 220 GHz opacity of ~ 0.1, at the 50° observing zenith angle), the longitude range of the Galactic plane covered at transit is ~ 15° - 230°: the whole of the second quadrant and most of the first and third quadrants. The Galactic plane in the longitude range ~ 35° - 210° transits at elevations > 60°, with longitudes 70° and 175° transiting at zenith. Therefore, the 2[nd] quadrant is the prime target area. Thus, the 492 GHz CI maps can cover the Perseus and the local arm regions. This includes the Taurus-Perseus-Auriga complex, the Cygnus region, the Cepheus and Polaris Flare regions. These regions have extended latitude coverage in the CfA CO(1-0) survey maps. Thus, the most well studied nearest low mass star-forming regions in Taurus, Perseus and Cepheus are within the reach of the Himalayan telescope. For 230 and 345 GHz, a much larger extent of the Galactic plane will be well observable due to higher transmission. In a later phase at a higher site, we can observe at lower elevations in the 461/492 GHz bands, including the Galactic center (~ 30 degrees elevation). Limited observations from Hanle in the best quartile weather conditions (220 GHz opacity of 0.05; PWV ~ 0.7 mm) will also allow similar access, in the initial phase. Such access is necessary for the diffuse gamma ray emission templates science (section 5.1.1.3). We use winter 1st quartile conditions for sensitivity estimates below. We also conservatively assume an antenna diameter of 3 m with a beam size of 50″ at 492 GHz, while noting that a 6 m or 10 -12 m diameter is the preferred choice for EHT and extragalactic observations (see below).

Simulations suggest that CI observations reaching $0.15 \times T_{CO(1-0)}$ will be able cover most of the CO-dark-$H_2$ (Papadopoulos et al, 2018), which we use as a guideline in setting the required sensitivity levels. Using the sensitivity of 0.3 K rms of the CfA CO survey in the 2nd quadrant, we need to target 0.045 K for the CI observations. The initial goal targets a CI survey at the same resolution as the CfA CO survey. For a 3 m antenna with 50″ beam, which has a hundred beams in one 8.5′ CfA CO survey beam, we need to reach 0.45 K rms per 50″ spectrum. In comparison, the HEAT survey observations of the (2-1) line have poorer sensitivities - 0.1 Kkm/s on 0.5 km/s channels on 2.2′ beam, or equivalently 0.2 K rms, translating to 0.05 K for 8.5′, similar to our target - but worse when scaled for the factor ~ 2 stronger (1-0) lines. The AST/RO observations are also less sensitive, 0.5 K rms on 1km/s channels and 1.8′ beam, translating to 0.1 K for 8.5′. Thus, our observations will reach more diffuse gas than found in the previous observations and better access the regions where the molecular cloud formation processes are active.

Using 1st quartile PWV of 0.9 mm for the winter months and an SSB receiver noise temperature of 135K (ALMA 2SB Rx) we estimate an average $T_{SYS, SSB}$ of ~ 1000 K over an observation. Assuming dual polarized receivers and position switching with half the time spent on a reference off-source position, we obtain the following noise estimate for 1s of observation with a 1 MHz channel width (0.6 km/s):

$$T_{rms,1s,1MHz} = \sqrt{2} \frac{T_{sys,SSB}}{\sqrt{0.5s*10^6 Hz}} / \sqrt{2} = 1.4 \ K \text{ or } 0.8 \text{ K km/s at 0.6 km/s resolution}$$

The time needed to reach 0.45K is 10 s. The number of 50″ beam spectra required to cover 90 sq. deg (1 quadrant; $\pm$ 0.5 deg. latitude) is 466000. The total time needed to complete this survey would be = 466000 x 10s = 4.7 x$10^6$ s = 1300 hrs, obtaining 4660 spectra (17 minutes each) at 8.5′ resolution. This translates to 76 days or 2.5 months, with 17 hrs of observing per day (70% uptime). The 1.5 months available in an year with this weather would allow 60% of the survey to be completed. Including time available at median weather, a CI survey of the 2nd quadrant can be completed in < 2 years. Extended coverage to a much larger galactic latitude range will be needed to map the Perseus and local arm regions fully, as they are nearby. Such an extension and mapping of a few selected additional clouds would require more time. Development of multi-pixel receivers and a future move to a higher site would ease the paths to these projects.

While a detailed observing strategy has not been worked out, the basic idea would be to observe each location of the Galactic plane at its transit to maximize sensitivity. Multiple observations, carried out in an on-the-fly mode, will be needed to accumulate the required sensitivity for each point which will also distribute the quality of the observing time over the map. Depending on spectral baseline stability, a single long off source measurement may be used for multiple target spectra, which presents a trade-off between low level correlated noise in the survey maps and time saving. Reaching the required signal to noise at full resolution will take much longer and will benefit from moving to a higher site (by a factor of ~ 2) and by upgrading to multi-pixel receivers in later phases. It will be desirable to conduct a survey of a smaller region at full resolution as a pilot. An initial coarse resolution pilot survey will also help refine observing strategy. We note that the size of the antenna, or equivalently the beam size, while deciding the best resolution in the survey, does not affect the survey speed. The full resolution data are retained and as we continue repeated mapping, the signal to noise ratio reaches interesting levels at progressively finer resolutions, until the limit set by the antenna beam size is attained. In regions with higher signal strength, high resolution maps will be available right from the starting stage. The detailed choices on specific regions to map, extended latitude coverage as opposed to the full accessible Galactic plane and the locations to target first for immediate impact, will be made as we develop the collaboration further.

For extragalactic observations, while we are not providing a specific sample of galaxies to observe, we show that there is sufficient sensitivity to carry out detection surveys and mapping observations of bright parts of nearby galaxies. We provide examples below to indicate feasibility. A more detailed observing case and plans, identifying specific regions and extents to observe in individual galaxies are yet to be defined. Taking 10 minutes and 10 km/s as representative integration time and spectral resolution, the expected noise level is $\sim 0.02$K under $1^{st}$ quartile weather.

Existing JCMT observations towards the Giant Molecular Association in M51 show a 0.6 K line peak, with line widths of $\sim 50$ km/s. Such a line is comfortably detected at $5\sigma$ in 25 seconds on 5 km/s channels. Thus, extended mapping of such regions is feasible. The 100 km/s wide line in M83 is 0.15 K. In M31, a single point JCMT observation of CI detected a 0.08 K line towards the D478 dark cloud (Israel, Tilanus & Bass, 1998) with 15 km/s linewidth. While this dark cloud does not represent the brighter GMC population in M31 based on CO observations, it is strong enough to be detected in 30 minutes of integration, implying more observable emission in other GMC regions. With the CO size of D478 being 6.5′ x 1.1′, such clouds will also allow CI mapping. In M33, JCMT found 0.2-0.5 K lines in four GMCs with $\sim 10$ km/s width (Wilson, 1997). The median size of dust traced clouds using SPIRE 250$\mu$m, SCUBA 450 & 850 $\mu$m data in M33 was found to be 105pc (Williams, Gear and Smith, 2019), which translates to 24″, which would lead to 0.05 and 0.125 K for the lines, requiring $5 - 30$ minutes for 5 sigma detections. A 6 m antenna with a 25″ beam would reduce this to $\sim 20$ s - 2 minutes. Thus, detection and mapping of these targets is feasible with a clear benefit from a 6 m antenna. Extended mapping will also benefit from a higher site. As further examples, a 0.25 K, 200 km/s wide line was detected towards the Circinus galaxy by APEX, and the 200 km/s wide line towards NGC 253 was $\sim 0.3$ K.

A majority of the available observations for more distant galaxies are from Herschel SPIRE FTS observations which do not resolve the lines, with a resolution of 1.2 GHz (720 km/s). The Herschel survey of ULIRGs (Israel, 2006) implies fluxes of $\sim 5$ Jy, which translates to 0.02 K lines with 200 km/s widths. The 1s rms for 200 km/s channels is 0.1 K. The 0.02 K line can be detected at $5\sigma$ in 10 minutes. Thus, a spectrally resolved survey is possible with a few hours of integration time per target. A 6 or 10-12 m antenna would again be preferred.

While we have restricted discussions to the proposed Himalayan telescope, the available observations outlined above also indicate viability of SMA observations towards selected extragalactic targets at high resolution reaching scales of individual GMCs. The stand-alone ACA component of ALMA, comparable to the SMA, has recently carried out a few such CI observations which again points to feasibility with the SMA. Similarly, within the Milky Way, the CI line intensities towards massive star-forming regions (Infrared Dark Clouds, IRDCs; Buether et al. 2014) indicate feasibility of higher resolution observations of similar regions with the SMA. The CO (4-3) line will be much more easily detected as it is stronger and the transmission is ~ 40% better. We note that such observations will require the use of the best 20% weather on Mauna Kea. All unresolved targets detected with a 3, 6 or 12 m Himalayan antenna are targets for higher resolution follow ups with the SMA. Further discussion of the feasibility and the science cases can be the topic of a separate document.

The sensitivity for EHT observations was discussed in section 5.2. We emphasize that a 6 m diameter for the antenna is a clearly preferred choice with a 10 – 12 m antenna offering very attractive prospects.

## 9. Future

The future of this initiative holds a lot promise, of which we provide a brief and incomplete outline here. The chief among the possibilities and the most ambitious with matching high impact would be the prospects for a partnership towards an EHT space VLBI station, already touched upon in the main document. ISRO possesses many of the capabilities needed but mm/sub-mm VLBI represents a quantum leap. The work and collaboration proposed here can pave the way for this development.

The most beneficial for the ground based effort would be the move to a higher site from where higher sub-mm and THz bands will be more readily accessible, due to better atmospheric transmission. The carbon mapping program can be carried out more efficiently and also be extended to the 2-1 line at 803 GHz. Towards this end, the prospective 5000 m site at Pologongka La can be characterized by either deploying the existing 220 GHz tipper or a newly developed tipper that would also extend measurements to other higher sub-mm bands. Deployment of a site testing Fourier Transform Spectrometer for THz measurements, developed at the CfA, can also be considered. The development of multi-pixel receivers is another avenue which will increase mapping speeds. Multi-pixel receiver development for all bands also presents synergies with future development paths for the SMA. With a move to a higher site, the development and deployment of receivers for higher bands would be natural follow ups.

The Himalayas arguably provide sites that are among the best for ground based observations in higher sub-mm and THz bands where atmospheric transmission is a critical factor, for the northern hemisphere in particular. Compared to Greenland, the lower latitude allows larger sky access. There are a number of unexplored THz spectral lines of interest, e.g.: the fine structure transitions of [NII] at $205\mu m$ (1.5 THz), [CII] at $158$ $\mu m$ (1.9 THz) considered the brightest and most important for ISM line cooling and high J CO transitions to trace very hot gas close to protostars. While airborne and space platforms are best suited to observations in these bands by avoiding the atmosphere, ground based facilities can support large instruments and would become increasingly important in the coming decades. Epoch of Reionization (EoR) and intensity mapping studies using the redshifted [CII] lines are also possible. This will require exceptional stability which should be first assessed through appropriate characterization studies.

Provision of multipixel, multiband continuum cameras (TES or KID technologies) will open up whole new fields of research for this observatory, e.g: Sunyaev-Zeldovich studies, dust continuum emission from Galactic star-forming regions and external galaxies. Studies of the CMB B-modes may also be appropriate for the sites and have been broached, with the prospects yet to be evaluated.

While other sites in the Ladakh region can be explored, the Pologongka La area may have the advantage of being close to the new Bilaspur-Leh railroad line, currently in a planning stage. The first phase of the location survey for this line is complete and the final location survey is on-going. This area includes higher locations without current road access. Higher sites in Ladakh, up to ~ 6000 m accessible by new roads being constructed may also be worth pursuing.

In closing, the proposed initiative opens a number of possibilities, extending beyond the immediate objectives. It represents the tip of an iceberg of much wider, mutually beneficial interaction between the Indian and U.S. astronomy communities.

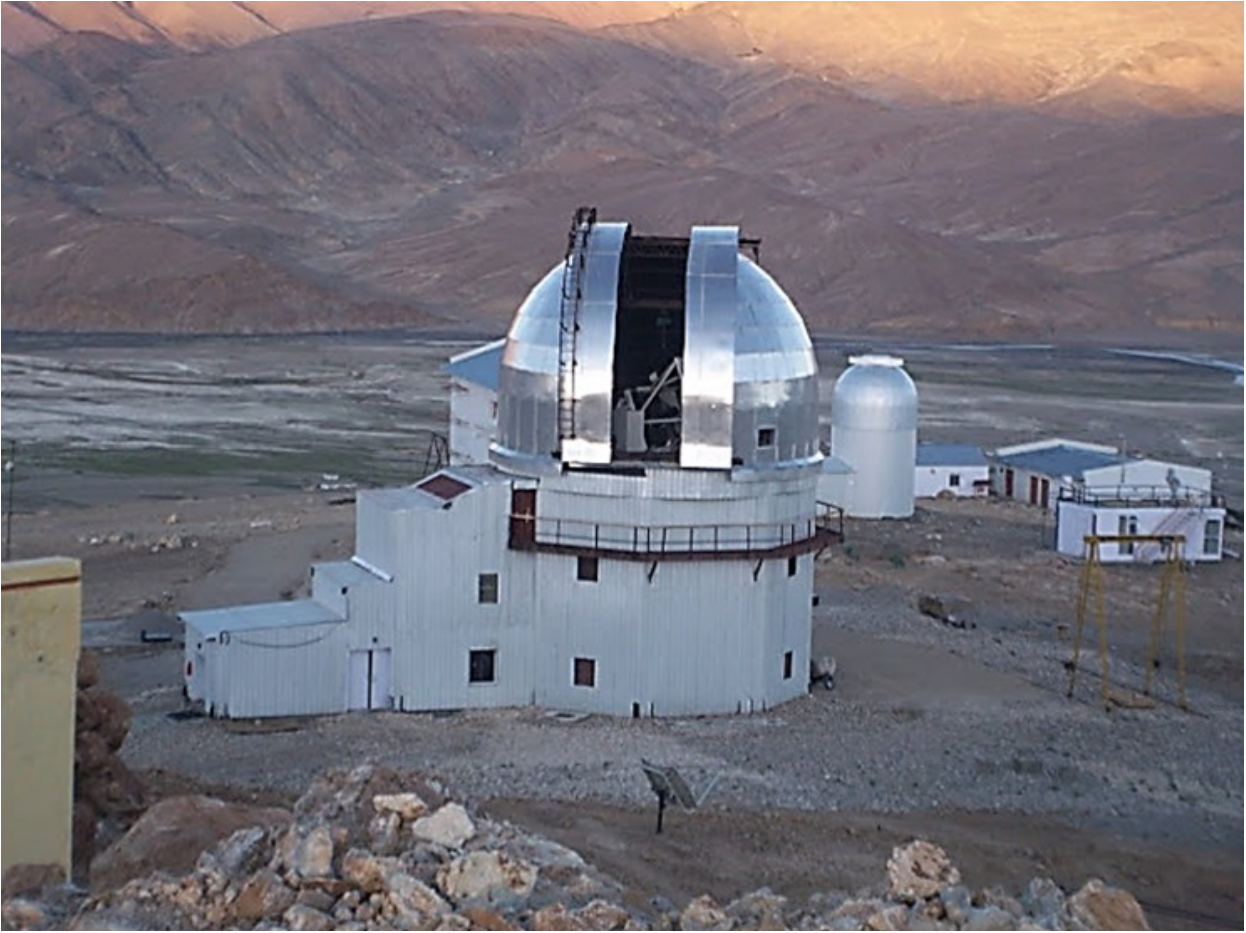

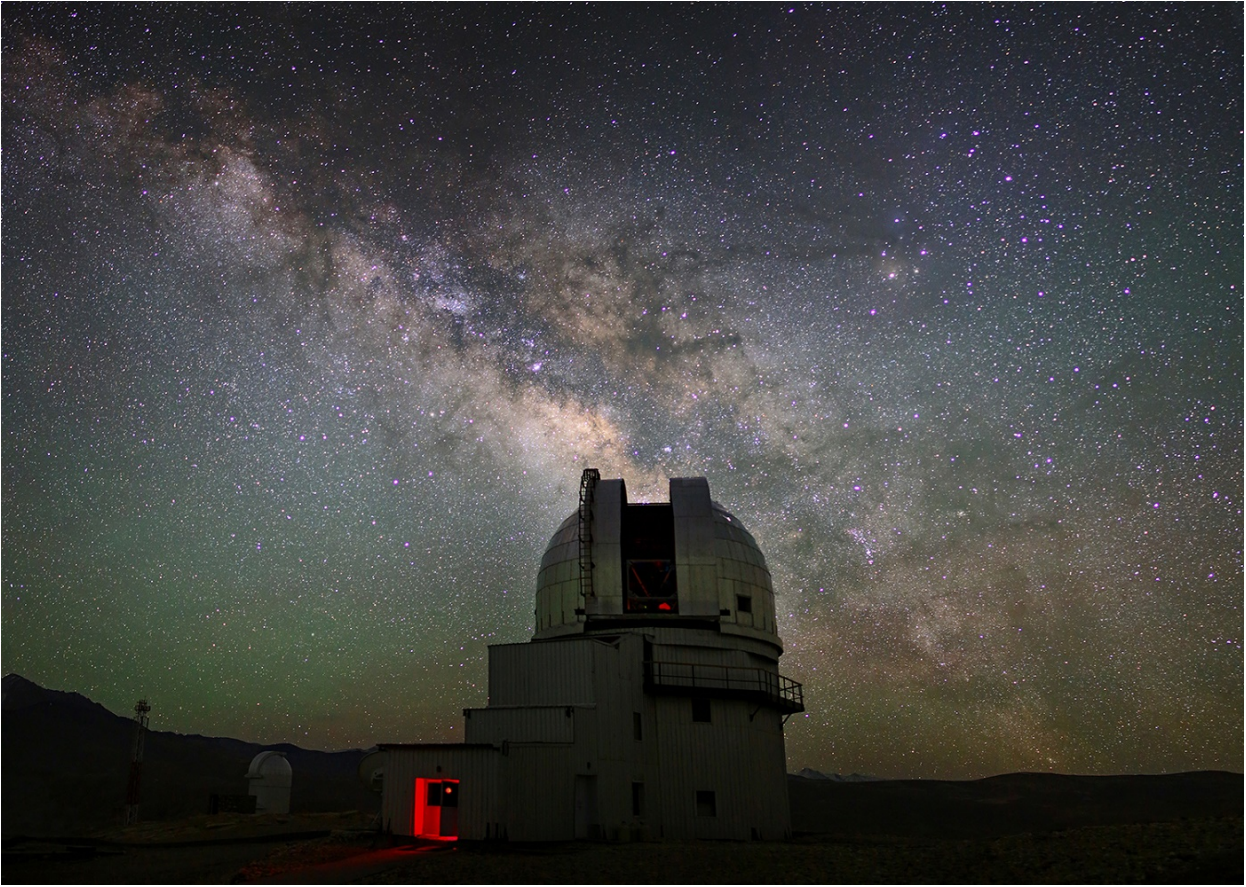

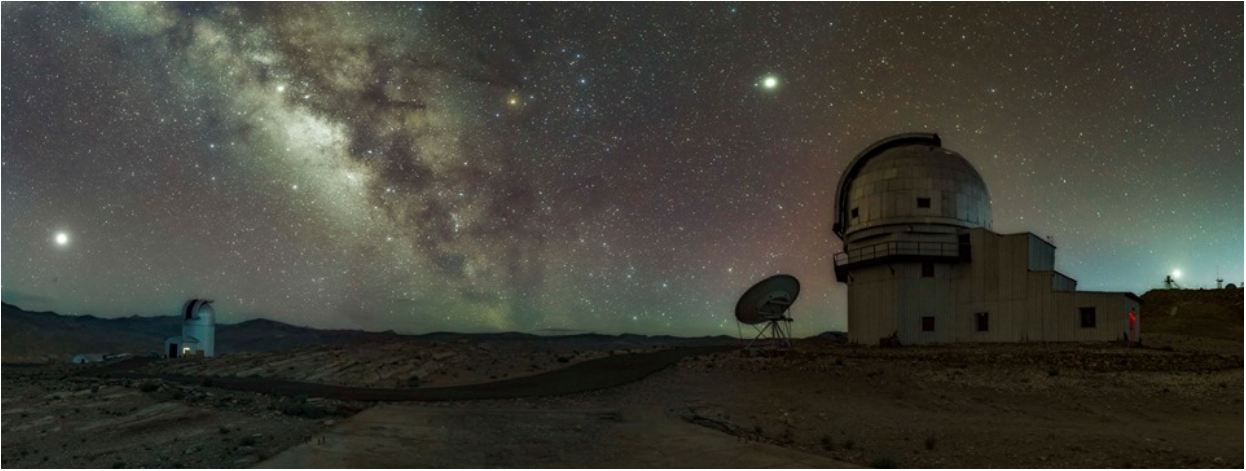